\documentclass[aps,prl,twocolumn,showpacs,floatfix,superscriptaddress]{revtex4-1}
\usepackage[pdftex]{graphicx}  % needed for figures
\usepackage{bm}        % for math
\usepackage{amssymb}   % for math
\usepackage{amsmath}
\usepackage[latin1]{inputenc}
\usepackage{natbib}
\usepackage{color}

\newcommand{\degree}{\ensuremath{^\circ}}

\begin{document}
\title{Plasmonic Brownian Ratchet}
\author{Paloma A. Huidobro}
\email[]{These two authors contribute equally to this work.}
\affiliation{Departamento de Fisica Teorica de la Materia Condensada and Condensed Matter Physics Center (IFIMAC), Universidad Autonoma de Madrid, 28049, Spain.}
\author{Sadao Ota}
\email[]{These two authors contribute equally to this work.}
\affiliation{National Science Foundation Nanoscale Science and Engineering Center, 3112 Etcheverry Hall, University of California at Berkeley, Berkeley, CA 94720, USA}
\author{Xiaodong Yang}
\affiliation{Department of Mechanical and Aerospace Engineering, Missouri University of Science and Technology, Rolla, Missouri 65409, USA.} 
\author{Xiaobo Yin}
\affiliation{National Science Foundation Nanoscale Science and Engineering Center, 3112 Etcheverry Hall, University of California at Berkeley, Berkeley, CA 94720, USA}
\affiliation{Materials Sciences Division, Lawrence Berkeley National Laboratory (LBNL), 1 Cyclotron Road, Berkeley, CA 94720, USA} 
\author{F.J. Garcia-Vidal}
\email[Corresponding author: ]{fj.garcia@uam.es}
\affiliation{Departamento de Fisica Teorica de la Materia Condensada and Condensed Matter Physics Center (IFIMAC), Universidad Autonoma de Madrid, 28049, Spain.}
\author{Xiang Zhang}
\email[Corresponding author: ]{xiang@berkeley.edu}
\affiliation{National Science Foundation Nanoscale Science and Engineering Center, 3112 Etcheverry Hall, University of California at Berkeley, Berkeley, CA 94720, USA}
\affiliation{Materials Sciences Division, Lawrence Berkeley National Laboratory (LBNL), 1 Cyclotron Road, Berkeley, CA 94720, USA}

\date{\today}

\begin{abstract}
Here we present a Brownian ratchet based on plasmonic interactions. By periodically turning on and off a laser beam that illuminates a periodic array of plasmonic nanostructures with broken spatial symmetry, the random thermal motion of a subwavelength dielectric bead is rectified into one direction. By means of the Molecular Dynamics technique we show a statistical directed drift in particle flow.
\end{abstract}

\maketitle

Photonic-based trapping and transport of nanoscopic objects is becoming increasingly relevant for microfluidic and lab-on-a-chip applications \cite{Erickson2011}. %Controlling objects in the nanoscale requires the trapping and transport of particles in a flow. 
Conventional optical tweezers, while being a cornerstone in biology and soft-condensed matter physics, provide trapping sizes intrinsically limited by diffraction \cite{Ashkin1987,Ashkin1987a}. Plasmonic nanostructures have attracted significant interest as a promising system to solve this issue \cite{Novotny1997,Chaumet2002}. Localized surface plasmons (LSP), collective excitation of free electrons in metals excited by light, are able to store electromagnetic (EM) energy at subwavelength scales \cite{Barnes2003,Maier2007}. This extremely enhanced EM energy in plasmonic antennas has enabled the trapping of Rayleigh dielectric particles and biological objects within subwavelength volumes \cite{Juan2011}. In principle, plasmonic antennas are designed to generate highly localized gradient forces for trapping the targeted objects into a single point fixed by the fabricated nanostructure \cite{Volpe2006,Grigorenko2008,Righini2009}. On the other hand, transport of these targets over long distances, which is important especially for on-chip sorting, has not been shown with this localized trapping. While large scattering forces were alternatively utilized to show the control of metallic particles on a flat substrate \cite{Wang2009,Cuche2012a}, the more powerful antenna-based tweezers have not been exploited for this purpose.

%Recent works have demonstrated the powerfulness of LSP-based nano-optical trapping of sub-micrometer polymer beads, polystere (PS) beads and even bacteria relying on gradient forces to attract either dielectric or biological targets to a single point \cite{Volpe2006,Grigorenko2008,Righini2009}. In general, these plasmonic nanotweezers are designed to suppress the particles' Brownian motion, which dominates the motion of nanoscopic objects in liquid \cite{Einstein1905}. Indeed, thermal noise usually represents an obstacle for the control of particle flow, as in Refs. \cite{Wang2009,Cuche2012a}, where dynamical control relying on large scattering forces was shown for metallic particles. 

Thermal noise induced by random collision of the liquid molecules \cite{Einstein1905}, is usually an obstacle for controlling particle flow at nanoscopic scales and is to be suppressed for any particle manipulation technology. By taking advantage of these thermal fluctuations instead suppressing them, this work addresses the issue of dielectric particle transport over long distances in plasmonic structures. Although the second law of thermodynamics states that temperature-governed fluctuating forces originate no net motion in the large scale, by driving an anisotropic system out of thermodynamic equilibrium work can be performed out of thermal noise. This is the working principle of the so-called Brownian ratchets \cite{Feynman1966,Hanggi2009}, which have been of fundamental interest in biology because they play a role in biological protein motors \cite{Takano2010}. These devices have also attracted great interest as the basic working principle of practical devices \cite{Faucheux1995,Astumian1997,Bader1999,VanOudenaarden1999,Zapata2009,DiLeonardo2010}. Moreover, mesoscopic Brownian motors, such as organic electronic ratchets and spin ratchets, have been experimentally demonstrated, showing that they are great candidates to power nanodevices and electronic circuits \cite{Song1998,Linke1999,Villegas2003,Khrapai2006,Costache2010}.   

In this Letter, we present a proof of principle demonstration of a light-driven nanoscale Brownian ratchet based on plasmonic interactions. % that utilizes the unavoidable thermal noise as a sufficiently large energy source for plasmonic devices with different functionalities. 
It makes use of plasmonic-based optical forces, that first enable the trapping of a subwavelength dielectric bead and then drives it a long distance displacement in a single device at room temperature. By means of an array of plasmonic structures with broken spatial symmetry, we generate a set of anisotropic traps for dielectric beads. This trapping potential is repeatedly excited by turning on and off a laser field, thus taking the system out of equilibrium and yielding a directed drift of particles into one direction. In addition, we demonstrate this mechanism by means of Molecular Dynamics (MD) simulations, showing the rectified Brownian motion of a dielectric bead in the absence of any external bias.

\begin{figure*}[hbt!]
\centering
	\includegraphics[width=.9\linewidth]{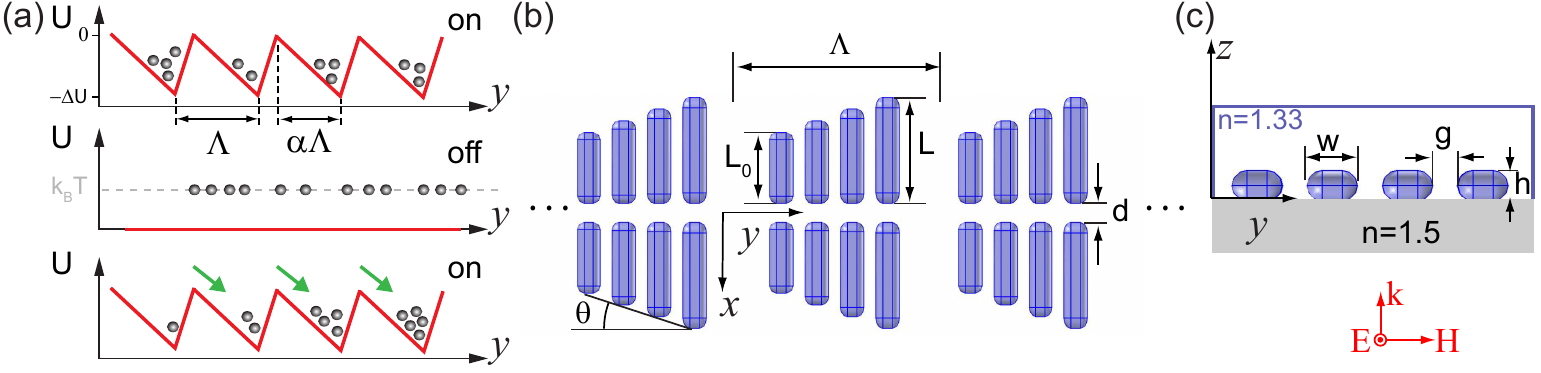}
	\caption{(a) By periodically turning on and off an external potential the Brownian diffusion of a dielectric particle is biased into one direction. (b) Top view of the plasmonic system: a periodic array of plasmonic structures formed by four optical dipole antennas. The geometrical asymmetry is characterized by angle $\theta$. (c) Side view of the periodic array unit cell: the plasmonic structure is placed on top of a glass substrate and embedded in water.}
\end{figure*}

The system we propose here utilizes an array of plasmonic nanostructures to bias the Brownian motion of subwavelength dielectric beads. The design of the plasmonic Brownian ratchet is based on (i) anisotropic trapping, generated by geometrical asymmetry, and (ii) a periodic modulation in time of the interaction, achieved by switching the illumination on and off in periodic cycles [see Fig. 1 (a)]. To fulfill these requirements we propose the periodic array of structures shown in Figs. 1 (b) and (c). The unit cell consists of four metallic dipole antennas (dielectric permittivity $\epsilon_m$) of gradually varying lengths and with deep subwavelength spacing \cite{Zhang2012}. Each dipole antenna has a subwavelength gap $d$ and is separated from its nearest neighbour by $g$. The nanorods composing the dipole antennas have width $\text{w}$, height $h$ and varying lengths: $L$ in one side of the group and $L_0=L-3(\text{w}+g)\tan(\theta)$ in another. The angle $\theta$ characterizes the asymmetry of the structure. The array of nanostructures, with period $\Lambda$, lies on top of a glass substrate and is immersed in water containing a solution of dielectric beads at room temperature. The system is illuminated with a normally incident plane wave (intensity $I_0$) with the electric field polarized along the dipole antennas axis ($E_x$), such that the near field within the gaps is efficiently enhanced and dielectric beads can be trapped in the EM hot spots. This way, we create a one-dimensional (1D) array of anisotropic nano-optical traps.

\begin{figure}[htb]
\centering
	\includegraphics[width=.95\linewidth]{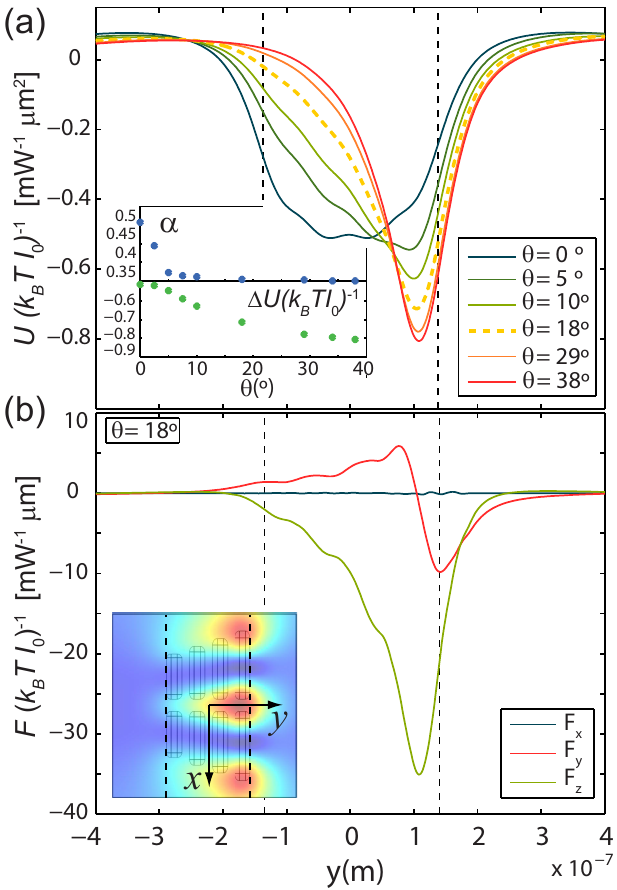}
	\caption{(a) Trapping potential experienced by a PS bead above the plasmonic nanostructure (four dipole antennas). Several values of the geometrical asymmetry are considered. Inset panels: evolution of the potential asymmetry parameter, $\alpha$ (upper), and depth of the potential well, $\Delta U$ (lower), as a function of $\theta$.  The dashed vertical lines point at the antennas boundaries for reference. (b) Forces experienced by the PS bead in the presence of a structure with $\theta=18\degree$. Inset panel: the norm of the electric field upon illumination at $\lambda=1.5\mu$m is plotted at $z=90$nm. The dipole antennas are designed to maximize the field enhancement at an infrared excitation wavelength: geometrical parameters $L=225$ nm, $d=40$ nm, $\text{w}=50$ nm, $g=25$ nm and $h=30$ nm.}
\end{figure}

We first characterize the plasmonic nanotweezers by considering a single array element (i.e., the four dipole antennas that compose an unit cell). A subwavelength dielectric bead in the presence of the plasmonic structure experiences a force that enables its trapping. This force is completely determined by the system EM fields, which are contained in the Maxwell stress tensor, $\tensor{\mathbf{T}}(\mathbf{r})$ \cite{Novotny2006}, and are calculated from simulations based on the finite element method \cite{comsol}. The integration of this tensor on an arbitrary surface $\delta V$ that encloses the particle yields the force exterted on it:
\begin{equation}
	\mathbf{F} =\int_{\delta V}\left\langle \tensor{\mathbf{T}} (\mathbf{r}) \right\rangle \cdot \mathbf{n}(\mathbf{r}) \text{d}a
\end{equation}
where $\mathbf{n}(\mathbf{r})$ is the normal to the integration surface. To obtain the force field \cite{Zhao2010}, the tensor $\tensor{\mathbf{T}}(\mathbf{r})$ is calculated for different positions of the particle's center (see Supplemental Material for more details). A line integration of the force yields the potential energy needed to move the particle from infinity to a position $\mathbf{r}$:
\begin{equation}
	U(\mathbf{r})=\int_{-\infty}^\mathbf{r}  \mathbf{F}(\mathbf{r'}) \cdot \text{d}\mathbf{r'}
\end{equation}
Figure 2 (a) shows the potential $U(y)$ obtained for a PS bead (refractive index $n=1.59$, radius $\sigma=50$ nm) for the plasmonic structure composed of four gold dipole antennas for several values of the angle $\theta$ [depicted in Fig. 1(b)]. We calculate the potential while a particle that is separated 10 nm to the plasmonic structure moves along the group's symmetry axis ($y$ direction, $x=0$). The potential is normalized to the thermal energy ($k_B T$) and the illumination intensity ($I_0$). The curves show that the asymmetric geometry yields the anisotropy in the trap. The simulated EM field for the structure with $\theta=18\degree$ is shown in the inset panel of Fig. 2(b). The field pattern reveals that the EM field is greatly enhanced in a hot spot located at the longest antenna's gap. The presence of the neighbouring antennas, with slightly different resonance frequencies due to their slightly different lengths, results in a continuous and asymmetric field profile along the $y$ direction. We have characterized the potential by means of its asymmetry, given by parameter $\alpha$ \footnote{The asymmetry parameter is defined as $\alpha=0.5-y_{min}/L$, where $y_{min}$ is the position of the potential minimum and $L=8\times10^{-7}$ m, with the antennas centered as shown in Fig. 2. Since we consider an individual potential well, we choose $L$ such that $U(-L/2)=U(L/2)$. The value of $\alpha$ ranges between 0 and 0.5. For a symmetric potential distribution, $\alpha=0.5$.}, and its depth, $\Delta U$ [both sketched in Fig. 1 (a)]. The evolution of these two parameters with $\theta$ is shown in the upper ($\alpha$) and lower ($\Delta U$) inset panels. The case $\theta=0\degree$ corresponds to a symmetric structure, which leads to a symmetric potential ($\alpha=0.5$). Increasing $\theta$ makes the potential more anisotropic ($\alpha$ decreases) by moving the hot spot towards the longest antenna's gap and, at the same time, it increases the depth of the potential well. For $\theta\approx18\degree$ [corresponding $U(y)$ plotted as a yellow dashed line in panel (a)], $\alpha$ saturates at $\approx 0.36$: for larger values of $\theta$, the dipole antennas are too different in length and they decouple to different resonance frequencies. Figure 2 (b) shows the force field that yields the potential in panel (a) for this case. As it can be seen from the figure, $F_y$ guides the particle towards the hot spot shown in the inset panel.%, as it is positive for $y$ values at the left of the hot spot, and negative when the bead is at the right of the hot spot. 
In the vertical direction, $F_z$ attracts the particle towards the structure, while in the $x$ direction $F_x\ll F_y,F_z$. These results show that the plasmonic structure with $\theta=18\degree$ is convenient for the design of a plasmonic ratchet as it gives a maximum anisotropy and values of the potential depth, $\Delta U \approx k_B T I_0\, [\text{mW}^{-1}\mu\text{m}^2]$, of the order of the thermal energy for illumination intensities $I_0 \approx 1$ mW$/\mu$m$^2$.

As required to design a Brownian ratchet, we now consider a periodic set of asymmetric plasmonic traps generated by arranging the plasmonic structures in a periodic array [see Fig. 1(b)]. Figure 3 shows the trapping potential calculated for a periodic array for different values of the period $\Lambda$. These curves stem from simulations of the plasmonic nanostructure subjected to periodic boundary conditions in the $y$ direction. Placing the plasmonic structures in an array gives rise to a coupling between them, yielding an oscillation of the potential depth with $\Lambda$ (see inset panel). The potential depth is maximum for $\Lambda\approx1$ $\mu$m, minimum for $\Lambda\approx1.4$ $\mu$m and approaches the $\Delta U $ value that corresponds to the single structure when the period is large enough. This oscillation is due to an interference effect in the far-field interaction between the plasmonic nanostructures. The far-field coupling also results in two local maxima in the potential $U(y)$ for values of $\Lambda\ge 900$ nm.

To demonstrate the Brownian ratchet operation, we have carried out MD simulations for a diffusive bead in the presence of the plasmonic structure while the trapping is periodically turned on and off. The dynamics of a particle of mass $m$ and radius $\sigma$ embedded in water at a constant temperature $T$ and subjected to an external force (the plasmonic force $F_{y}$) is governed by the following equation of motion (see Supplemental Material): 
\begin{equation}
	m\ddot{\mathbf{r}}=F_{y}-\gamma m\dot{\mathbf{r}}+ \sqrt{2 \gamma m k_B T} R(t)
	\label{eqLangevin}
\end{equation}
Here, the term $\gamma m\dot{\mathbf{r}}$ corresponds to the drag force of the particles in water, with $\gamma =6\pi\sigma\eta/m$. The interplay between the diffusion constant of the particle, $D=k_BT/m\gamma$, and the magnitude of the external forces dominates the dynamics. The last term in the equation is the Langevin expression that accounts for the Brownian motion of the particles in the fluid by means of a stochastic force: $R(t)$ is a delta-correlated gaussian function with zero mean \cite{Langevin1908}. 

 \begin{figure}[hb]
\centering
	\includegraphics[width=.95\linewidth]{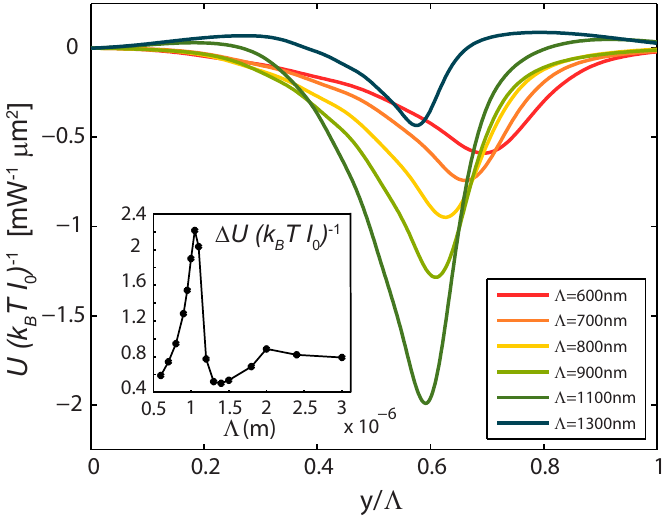}
	\caption{Unit cell of the trapping potential generated by a periodic array of plasmonic structures ($\theta=18\degree$). The anisotropic trapping potential is shown for different values of the period, $\Lambda$. Inset panel: evolution of the potential depth with the array period.}
\end{figure}

By solving Eq. \ref{eqLangevin} for $N$ realizations of the system at $T=300$ K we simulate the statistics of the system. For a proof of concept design of the ratchet, we take an array of period $\Lambda=800$ nm ($U(y)$ plotted as a yellow line in Fig. 3), as it gives a potential without any local maxima \footnote{Local maxima at the sides of the absolute minimum prevent particle trapping reducing the effectiveness of the device.}, and we simulate the dynamics in 1D. To ensure a highly-efficient trapping, we take an irradiance $I_0=75$mW$/\mu$m$^2$, although this device also works for lower laser powers. Initially, the bead is placed within a unit cell ($0\le y \le \Lambda$) of the anisotropic periodic potential. During time $t_{\text{on}}$, an external laser excites the set of optical traps. Despite the anisotropy, there are no large-scale gradients in the system and hence no statistical net motion of the particle is to be expected. When $t_{\text{on}}$ is long enough, the bead falls into the potential minimum as $\Delta U \gg k_B T$. Figure 4 (a) presents a histogram of $N$ realizations of the particle's position at $t=t_{\text{on}}$ showing a distribution around the minimum. During the ratchet operation, the system is driven out of equilibrium by a repeated modulation of the potential in time. When the potential is off, the particle is free to diffuse around the potential minimum. Due to the spatial anisotropy, the characteristic time, $\tau_{\text{F}}$, needed for the particle to travel to the neighbouring unit cell in the forward direction (distance $\alpha  \Lambda$), is shorter than the characteristic time, $\tau_{\text{B}}$, to travel backwards [distance $ (1-\alpha) \Lambda$]. If the potential is switched off for a time $t_{\text{off}}$ satisfying $\tau_{\text{B}}\gg t_{\text{off}}\gg\tau_{\text{F}}$, the probability of the particle moving forward is enhanced while that of moving backwards is suppressed. Thus, after one cycle, trapping the particle in the potential minimum located in the forward direction is more likely than in the backwards direction. Repeating the cycle causes a directed drift of the particle for a statistical ensemble. Figure 4 (b) shows the particle's positions after 16 on-off cycles (time $t=0.26$ s), demonstrating a statistical directed motion into the forward direction. Notice that the probability distribution around each trapping site is asymmetric, which may reflect the potential anisotropy. The time evolution of the probabilities of moving forward, $P_{\text{F}}=N_{\text{F}}/N$, and backward, $P_{\text{B}}=N_{\text{B}}/N$, ($N_{\text{F}}$ and $N_{\text{B}}$ being the number of realizations where the particle is at $y\ge \Lambda$ and $y\le 0$, respectively), is presented as an inset plot in panel (a). Together with $P_{\text{F}}$ and $P_{\text{B}}$ we show $P_0$, the probability of remaining in the initial unit cell. In the plot we show with circles the probabilities after 4, 8, 12 and 16 on-off cycles, revealing that $P_{\text{F}}$ increases with the number of on-off cycles up to $40\%$ whereas $P_{\text{B}}$ saturates at around $10\%$ for 4 cycles. Due to the statistical origin of the ratchet, a large number of repetitions is needed to observe an increase in $P_{\text{F}}$ and hence directed motion. In addition, we can estimate the evolution of the mean particle position, $\langle y \rangle$, with the number of cycles, $n$. From our results we have found that $\langle y \rangle=\langle y \rangle_0+n(\alpha'  \Lambda)P$, where $\langle y \rangle_0$ is the mean initial position and $P\approx 0.5 \exp(-\tau_{\text{F}}/t_{\text{off}})$ is the probability of moving forward in one cycle (see e.g. Ref. \cite{Bader1999}). Our simulation yield the values: $\alpha'=0.3$ (obtained from a fit and slightly different from the theoretical value, $\alpha=0.36$) and $P\approx0.13$. Thereby we can predict that in $\sim10^2$ cycles the bead's mean position will reach the fifth potential well in the forward direction.

\begin{figure}[ht]
\centering
	\includegraphics[width=.99\linewidth]{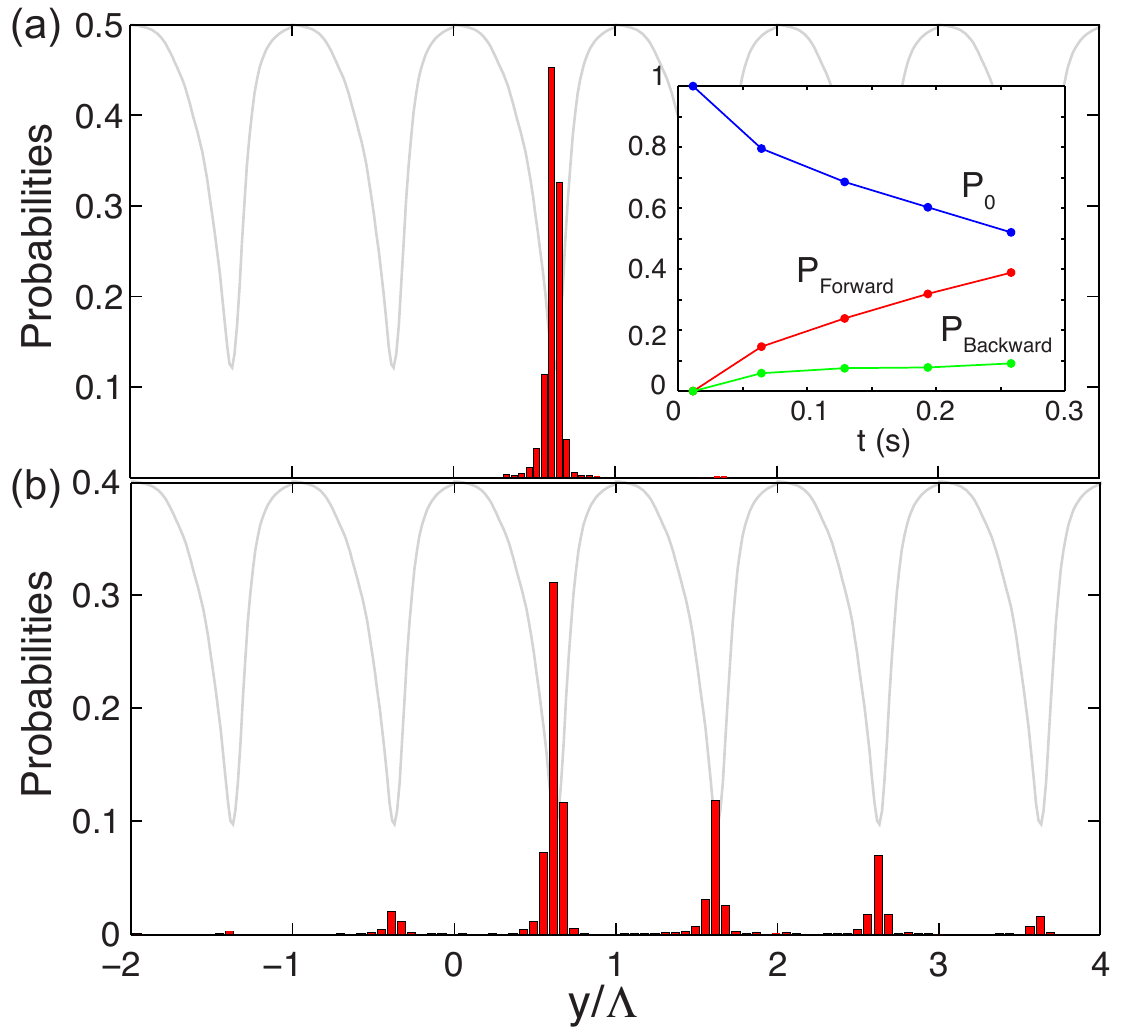}
	\caption{Plasmonic Brownian ratchet dynamics for $N=4000$ realizations of the system. One PS bead (radius $\sigma=50$ nm, density $\rho=1050$ $\text{kg}/\text{m}^3$) is embedded in water (viscous coefficient $\eta=1\times10^{-3}$ $\text{kg}/(\text{m}\, \text{s})$) at room temperature and subjected to the time-modulated plasmonic potential. The dynamics is diffusive ($\gamma \tau_{\sigma}\sim10^6$, $\tau_{\sigma}=\sigma^2/D$ being the  time for the bead to diffuse its own radius). (a) Initial situation: the particle probability distribution is centered at the trapping position of the nanotweezer ($0\le y \le \Lambda$) after a time $t_{\text{on}}=2\times10^{4}\tau_0$, with $\tau_0=\sqrt{m\sigma^2/k_BT}=5.7\times 10^{-7}$ s. Inset panel: time evolution of the probabilities for the particles to move forward ($P_{\text{F}}$), backward ($P_{\text{B}}$) or to remain in the initial unit cell ($P_0$). (b) Final situation: after $16$ on-off cycles ($t_{\text{off}}=8\cdot10^3{\tau_0}$), the probability for the particle to show a directed motion in the forward direction is $40\%$. The gray line represents the plasmonic potential in arbitrary units ($\Delta U=75k_BT$).}
\end{figure}

Finally, we discuss some issues related to the practical implementation of the plasmonic ratchet. It should be noted that the performance of the device can be affected by thermal-induced dynamics, as light coupling to LSPs in metallic particles leads to heat dissipation into the surroundings \cite{Baffou2013}. These effects should be minimized in an experiment by (i) reducing heat absorption in the metal by low illumination intensities, (ii) a careful design of the plasmonic structure (for instance, a substrate with a high thermal conductivity can work as a heat sink \cite{Wang11}) and (iii) considering thin fluidic cells \cite{Righini2009}. This last point has another advantage, as it prevents beads from escaping in the vertical direction when the illumination is turned off.   

To summarize, we have presented a proposal for transport of subwavelength dielectric particles in a plasmonic system that takes advantage of random thermal motion. This plasmonic Brownian ratchet only requires the periodic modulation in time of the large field enhancement generated in asymmetric plasmonic structures upon illumination. By means of a MD simulation, we have shown the statistical rectification of the Brownian motion of a sub-micrometer bead. The use of plasmonic structures allows to overcome the limitations of conventional techniques for light-based manipulation of particles, permitting to control the flow of subwavelength particles. This flexible device will find various applications including manipulation of colloidal and biological objects in micro- and nanofluidics.

We wish to thank T. Zentgraf for fruitful discussions. P.A.H. is grateful to T. Ruiz-Herrero for helping with MD simulations. This work was partially funded by the Spanish MINECO (No.MAT2011-28581-C02-01 and No. CSD2007-046-NanoLight.es). P.A.H. acknowledges FPU Grant AP2008-00021 from the Spanish Ministry of Education.

\bibliography{PBR}

\end{document}